\documentclass{article}

\usepackage[final, nonatbib]{neurips_2023_ml4ps}
\usepackage[sort&compress]{natbib} 

\usepackage[utf8]{inputenc} 
\usepackage[T1]{fontenc}    
\usepackage{hyperref}       
\usepackage{url}            
\usepackage{booktabs}       
\usepackage{amsfonts}       
\usepackage{nicefrac}       
\usepackage{microtype}      
\usepackage{xcolor}         
\usepackage{amsmath}
\usepackage{subcaption}
\usepackage{graphicx}

\title{Learning an Effective Evolution Equation for Particle-Mesh Simulations Across Cosmologies} 

\author{
\small{Nicolas Payot}\\
\small{Département de Physique, Université de Montréal}\\ \small{Mila - Quebec Artificial Intelligence Institute}\\
\small{Ciela - Montreal Institute for Astrophysical Data Analysis and Machine Learning}\\
\vspace{-0.3in}
\AND
\small{Pablo Lemos}\\
\small{Département de Physique, Université de Montréal}\\ \small{Mila - Quebec Artificial Intelligence Institute}\\
\small{Ciela - Montreal Institute for Astrophysical Data Analysis and Machine Learning}\\
\vspace{-0.3in}
\And
\small{Laurence Perreault-Levasseur}\\
\small{Département de Physique, Université de Montréal}\\ \small{Mila - Quebec Artificial Intelligence Institute}\\
\small{Ciela - Montreal Institute for Astrophysical Data Analysis and Machine Learning}\\
\small{Center for Computational Astrophysics, Flatiron Institute, NY}\\ 
\vspace{-0.3in}
\And
\small{Carolina Cuesta-Lazaro}\\
\small{The NSF AI Institute for Artificial Intelligence and Fundamental Interactions}\\ \small{Massachusetts Institute of Technology, Cambridge, MA}\\ 
\small{Center for Astrophysics | Harvard \& Smithsonian, Cambridge, MA}\\
\vspace{-0.3in}
\And
\small{Chirag Modi}\\
\small{Center for Computational Astrophysics, Flatiron Institute, NY}\\ 
\small{Center for Computational Mathematics, Flatiron Institute, NY}\\ 
\vspace{-0.3in}
\And
\small{Yashar Hezaveh}\\
\small{Département de Physique, Université de Montréal}\\ \small{Mila - Quebec Artificial Intelligence Institute}\\
\small{Ciela - Montreal Institute for Astrophysical Data Analysis and Machine Learning}\\
\small{Center for Computational Astrophysics, Flatiron Institute, NY}\\ 
\vspace{-0.3in}
}

\begin{document}

\maketitle

\begin{abstract}
 Particle-mesh simulations trade small-scale accuracy for speed compared to traditional, computationally expensive N-body codes in cosmological simulations. In this work, we show how a data-driven model could be used to learn an effective evolution equation for the particles, by correcting the errors of the particle-mesh potential incurred on small scales during simulations. We find that our learnt correction yields evolution equations that generalize well to new, unseen initial conditions and cosmologies. We further demonstrate that the resulting corrected maps can be used in a simulation-based inference framework to yield an unbiased inference of cosmological parameters. The model, a network implemented in Fourier space, is exclusively trained on the particle positions and velocities.  
  
\end{abstract}
\vspace{-0.50cm}
\section{Introduction}
\vspace{-0.10cm}
N-body simulations are a ubiquitous tool in astrophysics for modeling the dynamics of particles under the influence of their collective gravitational potential. While calculating interactions among a small number of particles can be relatively straightforward (e.g., with algorithms like Verlet integration), the computational burden escalates sharply with an $\mathcal{O}(n^2)$ time complexity, where $n$ denotes the number of particles in the simulation \citep{hockney_computer_2021}.

In cosmology, N-body simulations are employed to generate theoretical predictions for the large scale structure of the Universe by simulating the evolution of the dark matter distribution. These predictions are then used for comparison with observational data. With the advent of new generations of astronomical surveys probing increasingly large scales with unprecedented precision, there is a pressing need for N-body simulations that are both fast and precise on those scales. This has inspired the development of faster approximate methods, including solvers based on Lagrangian Perturbation Theory (LPT) \citep{LPT1,LPT2} and particle-mesh (PM) simulation-based approaches \citep{hockney_computer_2021}. Commonly used implementations of these techniques include 2LPT \citep{crocce_transients_2006}, FastPM \citep{feng_fastpm_2016}, and COLA \citep{tassev_solving_2013}.

Particle-mesh simulations work by binning the particles on a grid depending on their mass and position. The contribution of the particles to other grid points is calculated using a cloud-in-cell (CIC) interpolation scheme. Following this step, Poisson's equation is solved using Fast Fourier Transforms (FFTs) for the mass distribution on the grid. Once the potential is calculated, its inverse Fourier transform is obtained using another FFT operation. Forces acting on each particle can be interpolated using the obtained potential with another CIC. An update to particle positions and velocities are then computed and a time step is taken. Particle-mesh simulations are limited by either the number of particles (since assigning the particles to the grid mesh is of order $\mathcal{O}(n)$, where $n$ is the number of particle) or the potential calculation, which is of order $\mathcal{O}(m\log m)$, where $m$ is the number of grid nodes \citep{bodenheimer_numerical_2007, hockney_computer_2021}.

Unfortunately, these methods trade the small-scale accuracy of full N-body simulations for speed. To mitigate these limitations, PM simulations can be enhanced by techniques designed to correct the small-scale interactions, as done by  P3M method \citep{bodenheimer_numerical_2007}, or incorporating machine learning models. However, these methods typically focus on correcting small-scale without considering the specific cosmology being used (see e.g., \citet{he_learning_2019} for a neural network correction of LPT within a single cosmology), resulting in poor interpretability and questionable generalizability. As an alternative, the work of e.g.~\cite{2010MNRAS.405..143A, 10.1111/j.1365-2966.2011.19635.x} has focused on developing simple analytical schemes to rescale an N-body simulations at a particle or halo level to different cosmologies. However, such schemes cannot generate simulations with varying initial conditions, and assessing their accuracy is difficult beyond the first order quasi-linear regime. 

Inspired by the framework of the effective field theory of large scale structure \citep{EFTofLSS}, a more principled approach is to learn a correction to the evolution equation of the particles, that is to say, to learn an effective evolution equation capturing the error introduced by coarse-graining the gravitational potential on a mesh. This idea was initially explored in \citet{lanzieri_hybrid_2022}, however, their loss function had to include the ratio of the predicted and reference power spectra. Therefore, the accuracy of such correction for different, potentially more informative summary statistics is not guaranteed. In the present work, we show that it is possible to obtain an effective evolution equation that is robust to variations in cosmological parameters and initial conditions in a physically principled manner, by only imposing loss terms at the level of the position and velocities of the particles, effectively enforcing the conservation of global angular and linear momenta.

\vspace{-0.10cm}
\section{Methods}
\vspace{-0.10cm}
We used $\tt{JaxPM}$, a differential PM simulation package written in $\tt{JAX}$ \footnote{https://github.com/DifferentiableUniverseInitiative/JaxPM}. Following  \citet{lanzieri_hybrid_2022} and \citet{chatziloizos_deep_2022}, we adopted a fully connected neural network whose outputs represent the coefficients of a B-spline function with an order of $3$.  The network is an isotropic filter in Fourier space with sinusoidal activation functions \citep{zhumekenov_fourier_2019} to preserve translational and rotational symmetries.  Specifically, the network learns to correct the particle-mesh potential and subsequently applies these corrections to the potential in position space.

In $\tt{JaxPM}$, the particle-mesh solver consists of a set of ordinary differential equations (ODEs), enabling the back-propagation of the gradient to the initial conditions \citep{lanzieri_hybrid_2022}.
The ODE is:
\begin{align}
    \frac{dx}{da} &= \frac{v}{a^3E(a)}\\
    \frac{dv}{da} &= \frac{F_{\theta}(x, a, \Omega_m, \sigma_8)}{a^2E(a)} \, ,
\end{align}
with the force given by the equation (which includes the neural network correction to the potential):
\begin{equation}
    F_{\theta}(x, a, \Omega_m, \sigma_8) = \frac{3\Omega_m}{2}\nabla \left(\phi_{PM}(x) * \mathcal{F}^{-1}(1-f_{\theta}(a, |k|, \Omega_m, \sigma_8))\right) \, .
\end{equation}
Here, $x$ represents the positions of the particles, and $a$ corresponds to the cosmological scale factor. We use $\mathcal{F}^{-1}$ to denote inverse Fourier transform, and $f_{\theta}(a, |k|, \Omega_m, \sigma_8)$ to represent the neural network. In this study, we employ a total of $32$ knots, and the fully connected network consists of $5$ hidden layers, each with a size of $64$.

For training, we used simulations from the CAMELS suite \citep{villaescusa-navarro_camels_2021, ni_camels_2023, villaescusa-navarro_camels_2023}. We used IllustrisTNG dark matter only simulations from the LH set, which contains $1000$ simulations, each featuring distinct initial conditions and varying cosmological parameters \citep{2019ComAC...6....2N}. Specifically, $\Omega_m$ and $\sigma_8$ are sampled within the range $\left[0.1, 0.5\right]$ and $\left[0.6, 1\right]$, respectively, using a Latin Hypercube sampling method.
Each simulation consists of $256^3$ particles within a periodic comoving volume of $\left(25\,\text{Mpc}\,h^{-1}\right)^3$, spanning redshifts $z = 127$ to $z = 0$ and captured in $34$ snapshots. During training, the particle-mesh simulation is initialized at $z = 127$ and subsequently simulated until $z = 0$. The remaining $33$ snapshots are then used in the loss function, constructed with the $L2$ norm of the desired positions and velocities of the particles in the simulations:
\begin{equation}
    \mathcal{L} = \sum_i\sum_{j=0}^{33}\left( ||x_{ij}^{\rm nbody} -x_{ij}||_2^2 + \lambda||v_{ij}^{\rm nbody} -v_{ij}||_2^2\right) + \gamma\sum_i \beta_i^2 \, .
\end{equation}
Here, $x_{ij}$ and $v_{ij}$ represent the position and velocity of particle $i$ in snapshot $j$ and $\beta_i$ denotes the weights of the fully-connected neural network used for the correction, making the second term a simple $L2$ regularization. Two hyperparameters $\lambda$ and $\gamma$ adjust the contribution of the different losses and the impact of the regularization, respectively. 
\vspace{-0.10cm}
\section{Results}
\vspace{-0.10cm}
We explored with different hyperparameters in the loss function to find models with better performance. With $\lambda = 0$, which effectively removes the contribution of velocities to the loss, the velocities tended to be over-corrected, even with strong $\gamma$ regularization. With $\lambda = 1$, the $L2$ norm of the velocities tended to be bigger than that of the positions, causing the model to prioritize velocity corrections at the expense of positional accuracy. We found empirically that choosing $\lambda = 0.01$ and $\gamma = 1$ balanced the contributions of the three terms.

\begin{figure*}[t]
    \centering
        \centering
        \includegraphics[width=0.95\textwidth]{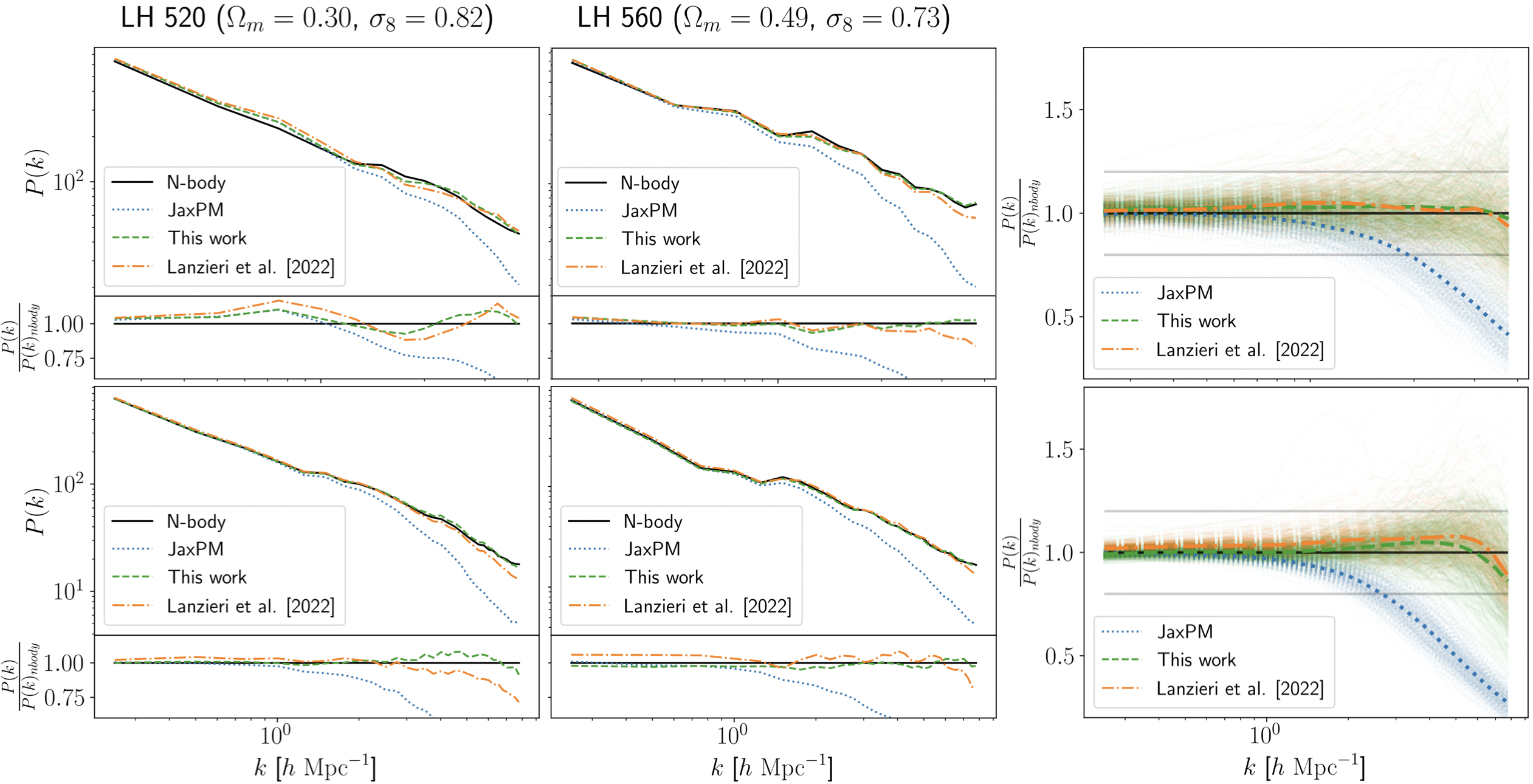}
     \caption{{\bf Top row:} Simulations with $32^3$ particles. {\bf Bottom row:} Simulations with $64^3$ particles. {\bf Left two columns:} Power spectra of two different simulations for CAMELS (black), JaxPM (blue), this work (green) and from \citep{lanzieri_hybrid_2022} (orange). {\bf Right column:} Fractional error of same three methods to the CAMELS simulation for $500$ simulations with cosmologies distinct from those seen during training. The thick lines represent the averages. Note that, as opposed to \citep{lanzieri_hybrid_2022}, the method presented here was not trained with the power spectrum explicitly in the loss. 
     }
    \label{fig:Pks}
\end{figure*}

After fixing these hyperparameters, we trained two models: one with $64^3$ particles and one with $32^3$. Both models were trained using a learning rate of $0.001$ on simulations ranging from LH100 to LH500, comprising a total of $400$ distinct cosmologies. They were trained respectively for $400$ and $500$ epochs on a single NVIDIA A100. The complete set of simulations could not be utilized because of limitations in storage capacity. Out of the remaining $600$ simulations, $100$ were used for validation, while the remaining $500$ were used for testing.

As shown in \citet{lanzieri_hybrid_2022} a learnt correction to the evolution equations can generalize well across different cosmologies, even when trained on a single set of cosmological parameters. We observed a similar result with our models; however, conditioning the neural network on the cosmological parameters performed even better, hence we only present the results of the latter strategy. Figure \ref{fig:Pks} illustrates how the correction can generalize across various cosmologies and even different initial conditions. As opposed to \citet{lanzieri_hybrid_2022}, the neural network does not have access to power spectra during training, making this a valuable tool for assessing the performance of our correction. This summary statistics is therefore a representative quantification of the performance of our correction, whereas as it is difficult to assess the effectiveness of the \citet{lanzieri_hybrid_2022} correction beyond the power spectrum (e.g. on other summary statistics).

Another key difference between this work and that of \citet{lanzieri_hybrid_2022} is that they train a different correction for every set of initial conditions, whereas this work trains a single model capable of evolving the simulation across any initial conditions. The distribution of power spectra across the entire range of simulations is depicted in the right column of Figure \ref{fig:Pks}. We find that $85\%$ of all simulations remain within $30\%$ of the N-body reference. The test simulations with the worse performances are the ones for which the power spectrum of the initial conditions is more than $10\%$ away from those seen during training. We therefore attribute this to the relatively small size of the training set, and believe that better performances could be achieve with more training examples.

\begin{figure}[t] 
   \centering
    \includegraphics[width=\textwidth]{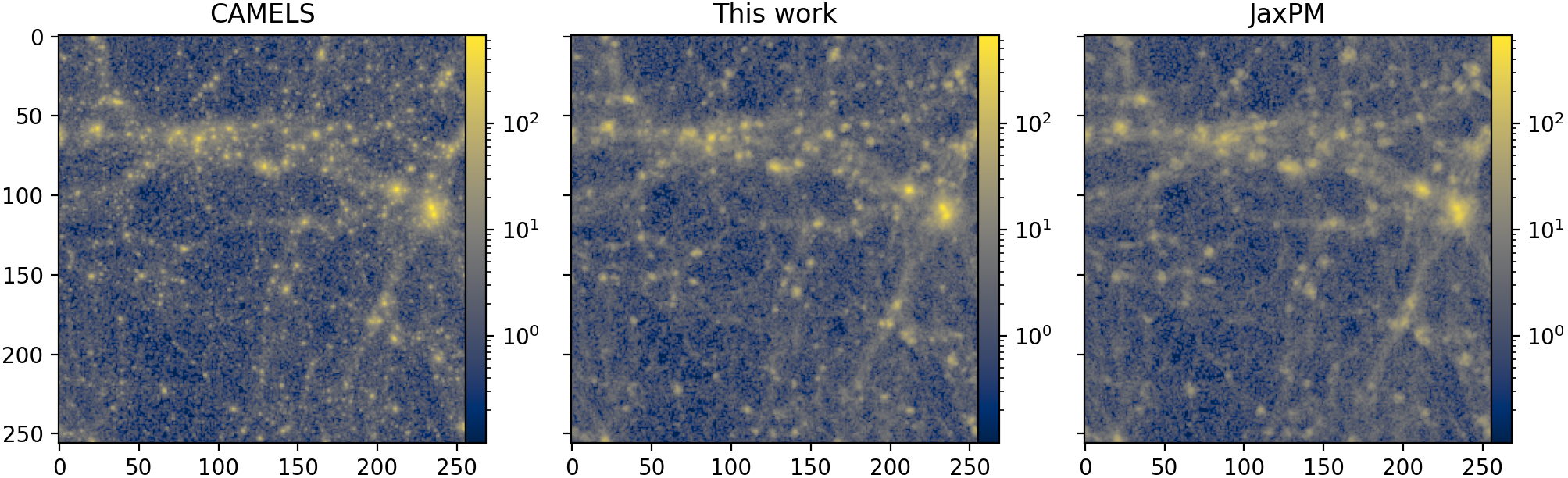}
    \caption{Comparison of z=0 density fields from the CAMELS simulation, our correction to $\tt{JaxPM}$, and the pure $\tt{JaxPM}$ simulation.}
    \label{fig:densityFields}
\end{figure}

\vspace{-0.15cm}
\section{Cosmological parameter inference}
\vspace{-0.15cm}
One of the key questions we wish to address is whether our learnt correction to $\tt{JaxPM}$ is close enough to a full N-body simulation that it could be used to perform unbiaised inference of cosmological parameters (specifically, $\Omega_M$ and $\sigma_8$) in a simulation-based inference (SBI) framework \citep{2020PNAS..11730055C}. As a first test of this, we use the power spectrum of our produced dark matter-only maps as the compressed statistics to train an Sequential Neural Posterior Estimator (SNPE) \citep{2022arXiv221004815D} using an ensemble of 5 masked auto-regressive flows (MAFs) to model the density, by making use of the $\tt{sbi}$ package\footnote{https://github.com/mackelab/sbi}. 

\begin{figure*}[t]
    \centering
        \includegraphics[width=\textwidth]{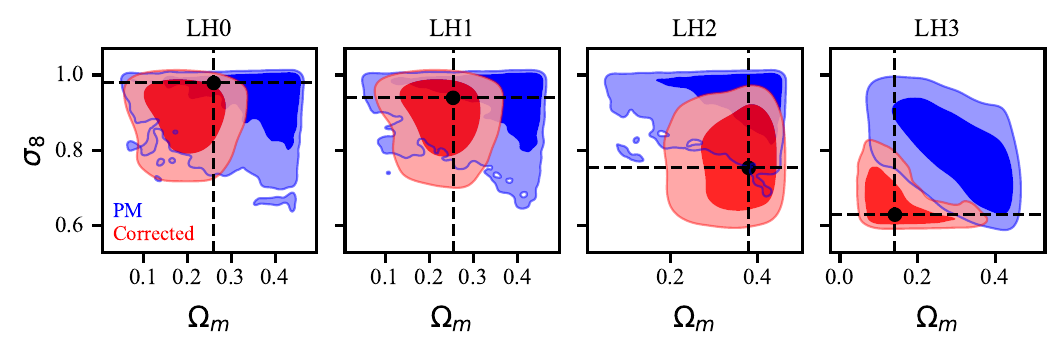}
    \caption{$1$ and $2\sigma$ constraints on $\Omega_m$ and $\sigma_8$ obtained in a SBI framework (with SNPE) using $\tt{JaxPM}$ as the simulator (blue contours) and the simulator in this work (red contours). The correction we propose significantly alleviates the bias otherwise induced by $\tt{JaxPM}$ in the inference.}
    \label{fig:sbi}
\end{figure*}

Once trained, we use a full N-body simulation obtained through the CAMELS dataset as the data and compute its power spectrum to infer $\Omega_M$ and $\sigma_8$. The results are presented as the red contours in Fig~\ref{fig:sbi}. As a point of comparison, we train another similar SNPE model using $\tt{JaxPM}$ as the simulator to obtain the power spectra to perform the same inference, and obtain the blue contours presented in Fig~\ref{fig:sbi}. As can be seen, the learnt correction significantly alleviates the biases that otherwise plague the inference. Alternatively, in order to alleviate this bias, one could truncate the power spectrum predicted by the uncorrected JaxPM to retain only scales at scales $k<1$ $h$Mpc$^{-1}$ to train the SNPE model, but, as shown in appendix A, the inference becomes proportionally less constraining, with error bars on recovered cosmological parameters roughly twice as large.

\vspace{-0.10cm}
\section{Conclusion}
\vspace{-0.15cm}
In this work, we have demonstrated how a model trained solely on particle positions and velocities can learn a data-diven correction to the equations of motion of dark matter particles in fast Particle-Mesh simulations, and effectively correct the power spectra for different initial conditions and cosmologies. Furthermore, employing these corrected simulations for cosmological parameter inference effectively mitigates bias arising from small scales inaccuracies in standard $\tt{JaxPM}$ simulations.

Through our choice of loss function, our work emphasizes the importance of imposing the preservation of known conserved quantities. We have found that this, together with learning a correction at the level of the evolution equation rather than directly in the data space, greatly improves robustness of the learnt simulator to different initial conditions and cosmologies. Moreover, this opens the door to more interpretability, as a scheme such as symbolic regression could be used to extract an analytical expression from the learnt correction.

\begin{ack}

This work is supported by the Simons Collaboration on “Learning the Universe". The Flatiron Institute is supported by the Simons Foundation. The work is in part supported by computational resources provided by Calcul Quebec and the Digital Research Alliance of Canada. Y.H. and L.P. acknowledge support from the Canada Research Chairs Program, the National Sciences and Engineering Council of Canada through grants RGPIN-2020- 05073 and 05102, and the Fonds de recherche du Québec through grants 2022-NC-301305 and 300397.

\end{ack}

\bibliographystyle{unsrtnat}
\bibliography{NeuripsPM}

\appendix
\section{Additional Figures}
\label{sec:A}
\begin{figure}[h]
    \centering
    \includegraphics[width=\textwidth]{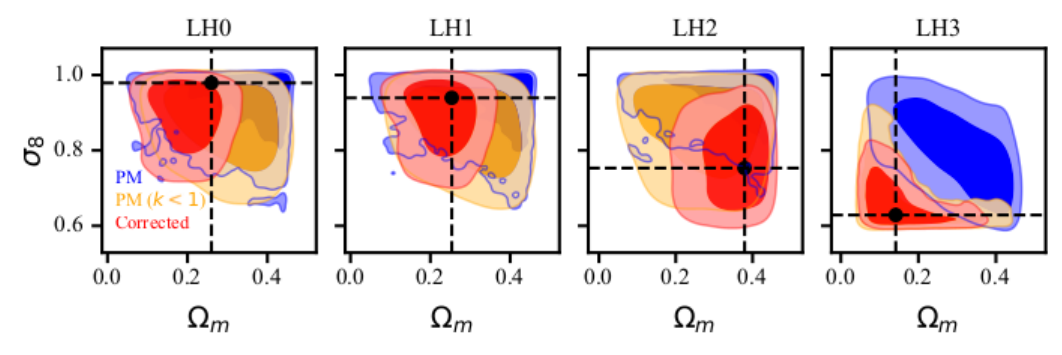}
    \caption{$1$ and $2\sigma$ constraints on $\Omega_m$ and $\sigma_8$ obtained in a SBI framework (with SNPE) using $\tt{JaxPM}$ as the simulator (blue contours), the $\tt{JaxPM}$ power spectra truncated to only retain scales with $k<1$~$h$Mpc$^{-1}$ (yellow contours), and the simulator in this work (red contours). The correction we propose significantly alleviates the bias otherwise induced by $\tt{JaxPM}$ in the inference, while truncating the $\tt{JaxPM}$ spectra significantly reduces the constraining power.}
    \label{fig:additionalSBI}
\end{figure}

\end{document}